\documentstyle[a4]{article}
\def\baselinestretch{1.2}

\newcommand{\eqref}[1]{$(\ref{#1})$}
\begin{document}
\begin{titlepage}
\vspace{55pt}

\begin{center}
  \begin{Large}
    \noindent{\bf Two-dimensional soliton cellular automaton}\\
    {\bf of deautonomized Toda-type}
  \end{Large}

\vspace{25pt}

\noindent
\begin{large}
A. {\sc Nagai}$^{1,2}$, T. {\sc Tokihiro}$^1$, J. {\sc Satsuma}$^1$,
R. {\sc Willox}$^{1,3}$ and K. {\sc Kajiwara}$^4$
\end{large}

\vspace{18pt}
\begin{small}
\noindent
${}^1${\it Graduate School of Mathematical Sciences, University of Tokyo,\\ 
Komaba 3-8-1, Meguro-ku, Tokyo 153, Japan}\\
${}^3${\it Dienst Tena, Vrije Universiteit Brussel, \\
Pleinlaan 2, 1050 Brussels, Belgium}\\
${}^4${\it Department of Electrical Engineering, Doshisha University,}\\
{\it Tanabe, Kyoto 610-03, Japan}\\
\end{small}

\vspace{18pt}
(Received \hspace{3cm})

\vfill
\hrule
\begin{abstract}
\noindent 
A deautonomized version of the two-dimensional Toda lattice equation
is presented. Its ultra-discrete analogue and soliton solutions
are also discussed.


\end{abstract}
\hrule
\end{center}
${}^2$E-mail: nagai@sat.t.u-tokyo.ac.jp

\vspace{1cm}
\noindent to appear in Phys. Lett. A.
\end{titlepage}
\section{Introduction}

The importance of discrete soliton equations has been recognized in
many fields such as mathematics, physics and engineering, owing to the
enormous development of computer science. Quite recently, two new
types of discretization have been proposed. One of these is to
unequalize lattice intervals~\cite{Hirotar}, which we call
``deautonomization'' in this paper. This idea comes from Newton's
interpolation formula~\cite{Milne} and now many applications in
numerical computation such as in the $\rho$-algorithm and in adaptive
numerical integration are expected. Another type of discretization is
to discretize dependent as well as independent variables, known as
``ultra-discretization''. One of the most important ultra-discrete
soliton systems is the so-called ``soliton cellular automaton'' or SCA
for short~\cite{Takahashi,Takahashi2}. In recent
papers~\cite{Tokihiro,Junta}, a general method to obtain the SCA from
discrete soliton equations was proposed. This procedure has been
applied to other soliton equations~\cite{Mori} as well.

The aim of this paper is to study a deautonomized version of the
two-dimensional(2D) Toda lattice equation and its ultra-discrete
analogue, which is considered as a deautonomized version of the 2D
Toda-type SCA~\cite{Mori}. In section 2, we review the result by Hirota
et al. on the 2D Toda lattice equation~\cite{Hirota2d} and derive its
deautonomized version. In section 3, we present an ultra-discrete
analogue of the deautonomized 2D Toda lattice equation, soliton
solutions for which are also discussed. Concluding remarks are given
in section 4.
 
\section{Deautonomization of the 2D Toda lattice equation}
We start with a discrete analogue of the 2D Toda lattice equation in
bilinear form~\cite{Hirota2d},
\begin{eqnarray}
&& (\Delta^+_{l}\Delta^-_{m}\tau(l,m,n))\tau(l,m,n)-
(\Delta^+_{l}\tau(l,m,n))\Delta^-_{m}\tau(l,m,n) \nonumber \\
&=& \tau(l,m-1,n+1)\tau(l+1,m,n-1) -
\tau(l+1,m-1,n)\tau(l,m,n), \ l,m,n \in {\bf Z},\label{toda}
\end{eqnarray}
where $\Delta^+_{l}$ and $\Delta^-_{m}$ stand for difference operators 
defined by
\begin{equation}
\Delta^+_{l} \tau =
\frac{\tau(l+1,m,n)-\tau(l,m,n)}{\delta},\ 
\Delta^-_{m} \tau =
\frac{\tau(l,m,n)-\tau(l,m-1,n)}{\epsilon}.
\label{diffop}
\end{equation}
Rewriting eq.~\eqref{toda} by using eq.~\eqref{diffop}, we obtain
\begin{eqnarray}
(1-\delta\epsilon)\tau(l+1,m-1,n)\tau(l,m,n)
 - \tau(l+1,m,n)\tau(l,m-1,n) \nonumber \\
 + \delta \epsilon \tau(l,m-1,n+1)\tau(l+1,m,n-1)
= 0, \label{toda2}
\end{eqnarray}
which possesses a particular solution expressed as~\cite{Hirota2d}
\begin{equation}
\tau(l,m,n) = \left|
\begin{array}{cccc}
f^{(1)}(l,m,n) & f^{(1)}(l,m,n+1) & \cdots & f^{(1)}(l,m,n+N-1)\\ 
f^{(2)}(l,m,n) & f^{(2)}(l,m,n+1) & \cdots & f^{(2)}(l,m,n+N-1)\\ 
\vdots & \vdots & & \vdots \\
f^{(N)}(l,m,n) & f^{(N)}(l,m,n+1) & \cdots & f^{(N)}(l,m,n+N-1)
\end{array}\right|. \label{tau}
\end{equation}
In eq.~\eqref{tau} each entry $f^{(i)}(l,m,n) \ (i=1,2,\cdots,N)$ 
satisfies the following dispersion relations;
\begin{equation}
\left\{ \begin{array}{rcl}
\Delta^+_{l}f^{(i)}(l,m,n)&=& f^{(i)}(l,m,n+1),\\
\Delta^-_{m}f^{(i)}(l,m,n)&=& - f^{(i)}(l,m,n-1).
\end{array} \right. \label{toda:disp}
\end{equation}
When we take
\begin{equation}
f^{(i)}(l,m,n) = p_i^n (1+\delta p_i)^l (1+\epsilon p_i^{-1})^{-m}
+ q_i^n (1+\delta q_i)^l (1+\epsilon q_i^{-1})^{-m}
\label{eq:rei1}
\end{equation}
as a solution for eq.~\eqref{toda:disp}, the
$\tau$-function~\eqref{tau} gives an $N$-soliton solution.

As opposed to the autonomous eq.~\eqref{toda}, where there is a
constant lattice interval in each direction, we now present a
deautonomized version, where lattice interval between neighboring
lattice points can be chosen freely. Let us replace $f^{(i)}(l,m,n)$
in eq.~\eqref{eq:rei1} by
\begin{eqnarray}
f^{(i)}(l_t,m_j,n) = 
p_i^n \frac{\displaystyle{\prod_{a}^{t-1} (1+\delta_a p_i)}}
{\displaystyle{\prod_{b}^{j-1} (1+\epsilon_b p_i^{-1})}}
+ q_i^n \frac{\displaystyle{\prod_{a}^{t-1} (1+\delta_a q_i)}}
{\displaystyle{\prod_{b}^{j-1} (1+\epsilon_b q_i^{-1})}},
\label{eq:rei2}\\
\delta_t = l_{t+1}-l_t, \ \epsilon_j = m_{j+1}-m_j, \ 
t,j \in \mbox{\bf Z},
\end{eqnarray}
where the products represent, for example,
\begin{equation}
\displaystyle{\prod_{a}^{t-1} (1+\delta_a p_i)} =
\left\{ \begin{array}{ll}
\displaystyle\prod_{a=0}^{t-1} (1+\delta_a p_i) & t \geq 1, \\
1 & t = 0, \\
\displaystyle\prod_{a=t}^{-1} (1+\delta_a p_i)^{-1} & t \leq -1.
\end{array}\right.
\end{equation}
Then the following dispersion relations hold;
\begin{equation}
\left\{ \begin{array}{rcccl}
\Delta^+_{l_t}f^{(i)}(l_t,m_j,n)&\equiv & 
\displaystyle\frac{f^{(i)}(l_{t+1},m_j,n)-f^{(i)}(l_t,m_j,n)}
{\delta_t} &=& f^{(i)}(l_t,m_j,n+1),\\
\Delta^-_{m_j}f^{(i)}(l_t,m_j,n)&\equiv & 
\displaystyle\frac{f^{(i)}(l_{t},m_j,n)-f^{(i)}(l_t,m_{j-1},n)}
{\epsilon_{j-1}} &=& - f^{(i)}(l_t,m_j,n-1).
\end{array}\right. \label{deauto:disp}
\end{equation}
By using the Pl\"ucker identity for determinants, we see that the 
$\tau$-function given by
\begin{equation}
\tau(l_t,m_j,n) = \left|
\begin{array}{cccc}
f^{(1)}(l_t,m_j,n) & f^{(1)}(l_t,m_j,n+1) & \cdots & f^{(1)}(l_t,m_j,n+N-1)\\ 
f^{(2)}(l_t,m_j,n) & f^{(2)}(l_t,m_j,n+1) & \cdots & f^{(2)}(l_t,m_j,n+N-1)\\ 
\vdots & \vdots & & \vdots \\
f^{(N)}(l_t,m_j,n) & f^{(N)}(l_t,m_j,n+1) & \cdots & f^{(N)}(l_t,m_j,n+N-1)
\end{array}\right|,
\end{equation}
with each $f^{(i)}$ solution of eq.~\eqref{deauto:disp},
satisfies a bilinear equation,
\begin{eqnarray}
(\Delta^+_{l_t}\Delta^-_{m_j}\tau(l_t,m_j,n))\tau(l_t,m_j,n)-
\Delta^+_{l_t}\tau(l_t,m_j,n)\Delta^-_{m_j}\tau(l_t,m_j,n) \nonumber \\
= \tau(l_t,m_{j-1},n+1)\tau(l_{t+1},m_{j},n-1) -
\tau(l_{t+1},m_{j-1},n)\tau(l_t,m_{j},n), \label{rtoda}
\end{eqnarray}
or equivalently,
\begin{eqnarray}
 (1-\delta_t\epsilon_{j-1})\tau(l_{t+1},m_{j-1},n)\tau(l_{t},m_{j},n)
 - \tau(l_{t+1},m_{j},n)\tau(l_{t},m_{j-1},n) \nonumber \\
 + \delta_t \epsilon_{j-1} \tau(l_{t},m_{j-1},n+1)\tau(l_{t+1},m_{j},n-1)
= 0, \label{2rtoda}
\end{eqnarray}
which is considered as a deautonomized 2D Toda lattice equation (in the 
sense explained above).

It is interesting to note that when we replace $f^{(i)}(l,m,n)$ by
\begin{eqnarray}
f^{(i)}(l_t,m_j,n_k) &=& 
\prod_{c}^{k-1} \left(\eta_c p_i\right) 
\frac{\displaystyle{\prod_{a}^{t-1} (1+\delta_a p_i)}}
{\displaystyle{\prod_{b}^{j-1} (1+\epsilon_b p_i^{-1})}}
+ \prod_{c}^{k-1} \left(\eta_c q_i\right) 
\frac{\displaystyle{\prod_{a}^{t-1} (1+\delta_a q_i)}}
{\displaystyle{\prod_{b}^{j-1} (1+\epsilon_b q_i^{-1})}},\\
\eta_c &=& n_{c+1} - n_c,
\end{eqnarray}
we can construct a new bilinear equation deautonomized with respect to
all independent variables. However, it reduce to eq.~\eqref{rtoda} by
simple dependent variable transformation.

\section{Ultra-discretization of the deautonomized 2D Toda lattice equation}

In this section, we present an ultra-discrete analogue of the
deautonomized Toda lattice eq.~\eqref{2rtoda} and we construct its
soliton solutions. Let us denote $\tau(l_t,m_j,n)$ as $\tau_{t,j,n}$
for simplicity. Equation~\eqref{2rtoda} is rewritten as
\begin{equation}
(1-\delta_t\epsilon_j)\tau_{{t+1},{j},n}\tau_{{t},{j+1},n}
- \tau_{{t+1},{j+1},n}\tau_{{t},{j},n} 
+ \delta_t \epsilon_{j} \tau_{{t},{j},n+1}\tau_{{t+1},{j+1},n-1}
= 0.
\label{rtoda2}
\end{equation}
When we introduce the new dependent variable $S_{t,j,n}$ as
\begin{equation}
\tau_{t,j,n} = \exp[S_{t,j,n}],
\end{equation}
eq.~\eqref{rtoda2} is equivalent to
\begin{eqnarray}
(1-\delta_t\epsilon_j) + \delta_t\epsilon_j \exp\left[
{\rm e}^{-\partial_n} (\Delta^+_n-\Delta^+_t)(\Delta^+_n-\Delta^+_j)
S_{t,j,n}\right]= \exp[\Delta^+_t\Delta^+_j S_{t,j,n}],
\label{Sijn} \\
\Delta^+_t = {\rm e}^{\partial_t} - 1,\
\Delta^+_j = {\rm e}^{\partial_j} - 1,\
\Delta^+_n = {\rm e}^{\partial_n} - 1.
\end{eqnarray}
Let us define a difference operator $\Delta'$ by
\begin{equation}
\Delta' = {\rm e}^{-\partial_n} 
(\Delta^+_n-\Delta^+_t)(\Delta^+_n-\Delta^+_j)
\end{equation}
for convenience sake.
Taking a logarithm and operating $\Delta'$ on both sides of 
eq.~\eqref{Sijn}, we have
\begin{eqnarray}
\Delta' \log(1-\delta_t\epsilon_j) + \Delta'\log
\left[ 1 + \frac{\delta_t\epsilon_j}{1-\delta_t\epsilon_j}\exp
(\Delta' S_{t,j,n})\right] = \Delta^+_t \Delta^+_j \Delta' S_{t,j,n}.
\label{rtodaS}
\end{eqnarray}
We finally take an ultra-discrete limit of eq.~\eqref{rtodaS}.
Setting
\begin{equation}
\Delta' S_{t,j,n} = \frac{v^\varepsilon_{t,j,n}}{\varepsilon}, \
\delta_t = {\rm e}^{-\theta_t/\varepsilon}, \
\epsilon_j = {\rm e}^{-\sigma_j/\varepsilon} \
(\theta_t, \sigma_j \in {\bf Z}_{\geq 0}),
\end{equation}
we obtain 
\begin{eqnarray}
\Delta^+_t \Delta^+_j v_{t,j,n} = 
{\rm e}^{-\partial_n} (\Delta^+_n - \Delta^+_t)(\Delta^+_n - \Delta^+_j)
F(v_{t,j,n}-\theta_t-\sigma_j), \label{rutoda}\\
F(x) = \max(0,x)
\end{eqnarray}
in the limit $\varepsilon \rightarrow +0$. We have rewritten
$\displaystyle{\lim_{\varepsilon \rightarrow +0}}
v^\varepsilon_{t,j,n}$ as $v_{t,j,n}$. Equation~\eqref{rutoda} is
considered as an ultra-discrete analogue of the deautonomized 2D Toda
lattice equation.

Next, we discuss soliton solutions for eq.~\eqref{rutoda}. It is
natural to consider that soliton solutions for eq.~\eqref{rutoda} are
obtained by an ultra-discretization of those for eq.~\eqref{rtoda}.
First of all, a one-soliton solution for eq.~\eqref{rtoda} is given by
\begin{eqnarray}
\tau_{t,j,n} &=& 1 + \eta_1,\\ 
\eta_1 &=& \alpha_1 \left(\frac{p_1}{q_1}\right)^n
\prod_{a}^{t-1} \frac{1+\delta_a p_1}{1+\delta_a q_1}
\prod_{b}^{j-1} \frac{1+\epsilon_b q_1^{-1}}{1+\epsilon_b p_1^{-1}}.
\end{eqnarray}
Introducing new parameters and variable as
\begin{eqnarray}
{\rm e}^{P_1/\varepsilon} = p_1, \ {\rm e}^{Q_1/\varepsilon} = q_1, 
\ {\rm e}^{A_1/\varepsilon} = \alpha_1, \\
{\rm e}^{-\theta_t/\varepsilon} = \delta_t , \
{\rm e}^{-\sigma_j/\varepsilon} = \epsilon_j, \\
\rho_{t,j,n}^\varepsilon = \varepsilon \log \tau_{t,j,n}
\end{eqnarray}
and taking the limit $\varepsilon \rightarrow +0$, we obtain
\begin{eqnarray}
\lim_{\varepsilon\rightarrow +0} \rho_{t,j,n}^\varepsilon \equiv
\rho_{t,j,n} &=& \max(0,K_1) \\
K_1 &=& A_1 + n (P_1-Q_1) + 
\sum_{a}^{t-1}\{F(P_1-\theta_a)-F(Q_1-\theta_a)\} \nonumber \\
&& + \sum_{b}^{j-1}\{F(-Q_1-\sigma_b)-F(-P_1-\sigma_b)\},\\
\displaystyle{\sum_{a}^{t-1}} &=&
\left\{ \begin{array}{ll}
\displaystyle\sum_{a=0}^{t-1} & t \geq 1, \\
0 & t = 0, \\
-\displaystyle\sum_{a=t}^{-1} & t \leq -1. 
\end{array}\right. 
\end{eqnarray}
A one-soliton solution for eq.~\eqref{rutoda} is given by
\begin{eqnarray}
v_{t,j,n} &=& \rho_{t+1,j+1,n-1}+\rho_{t,j,n+1}-\rho_{t+1,j,n}
-\rho_{t,j+1,n} \\
&=& \max\left(0,K_1-P_1+Q_1+F(P_1-\theta_t)-F(Q_1-\theta_t)
-F(-P_1-\sigma_j)+F(-Q_1-\sigma_j)\right) \nonumber \\
&& + \max(0,K_1+P_1-Q_1) 
- \max(0,K_1+F(P_1-\theta_t)-F(Q_1-\theta_t)) \nonumber \\
&& - \max(0,K_1-F(-P_1-\sigma_j)+F(-Q_1-\sigma_j)).
\end{eqnarray}

Secondly, we construct a two-soliton solution. Equation~\eqref{rtoda}
admits a two-soliton solution,
\begin{eqnarray}
\tau_{t,j,n} &=& 1 + \eta_1 + \eta_2 + \theta_{12}\eta_1\eta_2,\\ 
\eta_i &=& \alpha_i \left(\frac{p_i}{q_i}\right)^n
\prod_{a}^{t-1} \frac{1+\delta_a p_i}{1+\delta_a q_i}
\prod_{b}^{j-1} \frac{1+\epsilon_b q_i^{-1}}{1+\epsilon_b p_i^{-1}} 
\quad (i=1,2), \\
\theta_{12} &=& \frac{(p_2-p_1)(q_1-q_2)}{(q_1-p_2)(q_2-p_1)}.
\end{eqnarray}
In order to take an ultra-discrete limit for the above solution,
we suppose without loss of generality that the inequality,
\begin{equation}
0 < p_1 < p_2 < q_2 < q_1
\end{equation}
holds. Introducing new parameters and variable as
\begin{eqnarray}
&& {\rm e}^{P_i/\varepsilon}=p_i, \ {\rm e}^{Q_i/\varepsilon}=q_i, \
{\rm e}^{A_i/\varepsilon} = \alpha_i \
(i=1,2,\ P_1 < P_2 < Q_2 < Q_1), \\
&& \rho_{t,j,n}^\varepsilon = \varepsilon \log \tau_{t,j,n}
\end{eqnarray}
and taking the limit of small $\varepsilon$, we obtain 
\begin{eqnarray}
\lim_{\varepsilon\rightarrow +0} \rho_{t,j,n}^\varepsilon 
\equiv \rho_{t,j,n} &=& \max(0,K_1,K_2,K_1+K_2+P_2-Q_2),\\
K_i &=& A_i + n (P_i-Q_i) + 
\sum_{a}^{t-1}\{F(P_i-\theta_a)-F(Q_i-\theta_a)\} \nonumber \\
&& + \sum_{b}^{j-1}\{F(-Q_i-\sigma_b)-F(-P_i-\sigma_b)\}
\quad (i=1,2).
\end{eqnarray}

Finally, an $N$-soliton solution can also be constructed, under the
assumption that the inequality,
\begin{equation}
0 < p_1 < p_2 < \cdots < p_N < q_N < \cdots < q_2 < q_1 
\end{equation}
holds. Through the same limiting procedure, we have
\begin{equation}
\rho_{t,j,n} = \max_{\mu=0,1} \left(
\sum_{i=1}^N \mu_i K_i + 
\sum_{1\leq i<i'\leq N} \mu_i \mu_{i'} (P_{i'} - Q_{i'})\right),
\label{Nsol:ultra}
\end{equation}
where $\displaystyle{\max_{\mu=0,1}}$ is the maximization over all
possible combinations of $\mu_1 =0,1, \ \mu_2 =0,1, \cdots, \ \mu_N
=0,1$. Since the bounded and uniform convergence of the dependent
variable $\rho^\varepsilon_{t,j,n}$ as $\varepsilon \rightarrow +0$
is the cornerstone of the ultra-discretization, it should be clear
that eq.~\eqref{Nsol:ultra}, which is the ultra-discretization of 
the $N$-soliton solution for the discrete 2D Toda lattice equation,
is an $N$-soliton solution for the ultra-discrete equation.

We show one- and two-soliton solutions for eq.~\eqref{rutoda} with
$\sigma_j, \theta_t$ constant in Figures 1,3 and with $\sigma_j,
\theta_t$ chosen randomly in Figures 2,4. It should be noted that the
solutions with $\sigma_j, \theta_t$ constant are equivalent to those
discussed in~\cite{Mori}. We see that arbitrariness of $\sigma_j$ and
$\theta_t$ affects the values of the dependent variable $v_{n,t,j}$.
\clearpage
\def\baselinestretch{0.9}
\twocolumn
\begin{figure}[htbp]
\begin{center}
\begin{minipage}[t]{7cm}
\small
{\baselineskip 8pt
\phantom{n}\verb+              t=0             +\\
\phantom{n}\verb+ ..........................451 +\\
\phantom{n}\verb+ .......................154... +\\
\phantom{n}\verb+ .....................253..... +\\
\phantom{n}\verb+ ...................352....... +\\
\phantom{n}\verb+ .................451......... +\\
\phantom{n}\verb+ ..............154............ +\\
{n}\verb+ ............253.............. +\\
\phantom{n}\verb+ ..........352................ +\\
\phantom{n}\verb+ ........451.................. +\\
\phantom{n}\verb+ .....154..................... +\\
\phantom{n}\verb+ ...253....................... +\\
\phantom{n}\verb+ .352......................... +\\
\phantom{n}\verb+               j              +\\
}
\end{minipage}
\begin{minipage}[t]{7cm}
\small
{\baselineskip 8pt
\phantom{n}\verb+              t=5             +\\
\phantom{n}\verb+ ........................154.. +\\
\phantom{n}\verb+ ......................253.... +\\
\phantom{n}\verb+ ....................352...... +\\
\phantom{n}\verb+ ..................451........ +\\
\phantom{n}\verb+ ...............154........... +\\
\phantom{n}\verb+ .............253............. +\\
{n}\verb+ ...........352............... +\\
\phantom{n}\verb+ .........451................. +\\
\phantom{n}\verb+ ......154.................... +\\
\phantom{n}\verb+ ....253...................... +\\
\phantom{n}\verb+ ..352........................ +\\
\phantom{n}\verb+ 451.......................... +\\
\phantom{n}\verb+               j              +\\
}
\end{minipage}
\begin{minipage}[t]{7cm}
\small
{\baselineskip 8pt
\phantom{n}\verb+              t=10             +\\
\phantom{n}\verb+ .......................253... +\\
\phantom{n}\verb+ .....................352..... +\\
\phantom{n}\verb+ ...................451....... +\\
\phantom{n}\verb+ ................154.......... +\\
\phantom{n}\verb+ ..............253............ +\\
\phantom{n}\verb+ ............352.............. +\\
{n}\verb+ ..........451................ +\\
\phantom{n}\verb+ .......154................... +\\
\phantom{n}\verb+ .....253..................... +\\
\phantom{n}\verb+ ...352....................... +\\
\phantom{n}\verb+ .451......................... +\\
\phantom{n}\verb+ 4............................ +\\
\phantom{n}\verb+               j              +\\
}
\end{minipage}
\end{center}
\end{figure}
\pagebreak
\begin{figure}[htbp]
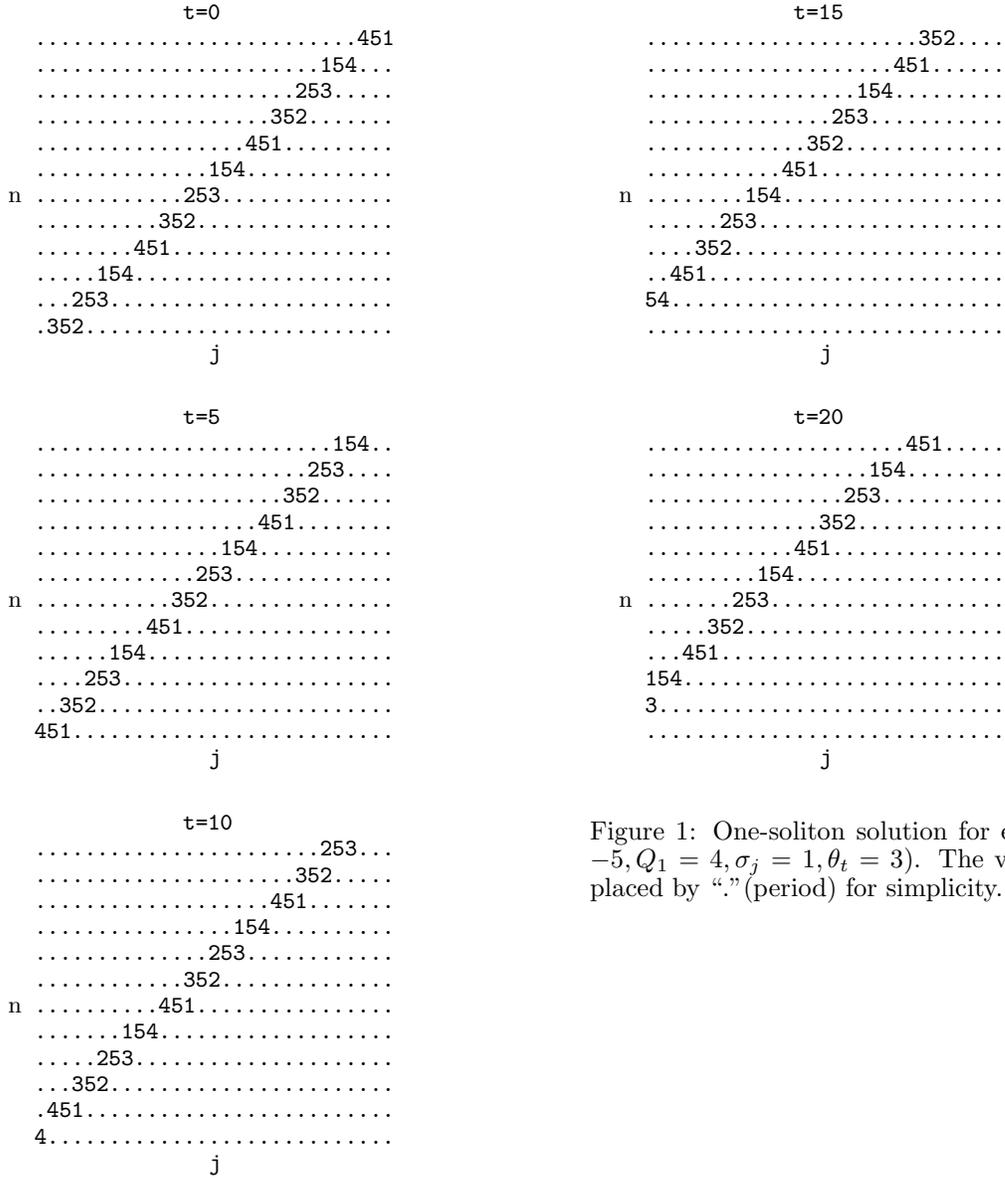

\begin{center}
\begin{minipage}[t]{7cm}
\small
{\baselineskip 8pt
\phantom{n}\verb+              t=15             +\\
\phantom{n}\verb+ ......................352.... +\\
\phantom{n}\verb+ ....................451...... +\\
\phantom{n}\verb+ .................154......... +\\
\phantom{n}\verb+ ...............253........... +\\
\phantom{n}\verb+ .............352............. +\\
\phantom{n}\verb+ ...........451............... +\\
{n}\verb+ ........154.................. +\\
\phantom{n}\verb+ ......253.................... +\\
\phantom{n}\verb+ ....352...................... +\\
\phantom{n}\verb+ ..451........................ +\\
\phantom{n}\verb+ 54........................... +\\
\phantom{n}\verb+ ............................. +\\
\phantom{n}\verb+               j              +\\
}
\end{minipage}
\begin{minipage}[t]{7cm}
\small
{\baselineskip 8pt
\phantom{n}\verb+              t=20             +\\
\phantom{n}\verb+ .....................451..... +\\
\phantom{n}\verb+ ..................154........ +\\
\phantom{n}\verb+ ................253.......... +\\
\phantom{n}\verb+ ..............352............ +\\
\phantom{n}\verb+ ............451.............. +\\
\phantom{n}\verb+ .........154................. +\\
{n}\verb+ .......253................... +\\
\phantom{n}\verb+ .....352..................... +\\
\phantom{n}\verb+ ...451....................... +\\
\phantom{n}\verb+ 154.......................... +\\
\phantom{n}\verb+ 3............................ +\\
\phantom{n}\verb+ ............................. +\\
\phantom{n}\verb+               j              +\\
}
\end{minipage}
\caption{One-soliton solution for eq.~\eqref{rutoda} ($P_1 = -5, 
Q_1 = 4, \sigma_j = 1, \theta_t = 3$).
The value ``0'' is replaced by ``.''(period) for simplicity.}
\end{center}
\end{figure}
\clearpage

\begin{figure}[htbp]
\begin{center}
\begin{minipage}[t]{7cm}
\small
{\baselineskip 8pt
\phantom{n}\verb+              t=0             +\\
\phantom{n}\verb+ ......................144.... +\\
\phantom{n}\verb+ ....................154...... +\\
\phantom{n}\verb+ ..................143........ +\\
\phantom{n}\verb+ ................144.......... +\\
\phantom{n}\verb+ ..............154............ +\\
\phantom{n}\verb+ ............143.............. +\\
{n}\verb+ ..........154................ +\\
\phantom{n}\verb+ .........43.................. +\\
\phantom{n}\verb+ .......44.................... +\\
\phantom{n}\verb+ .....44...................... +\\
\phantom{n}\verb+ ...451....................... +\\
\phantom{n}\verb+ .45.......................... +\\
\phantom{n}\verb+               j              +\\
}
\end{minipage}
\begin{minipage}[t]{7cm}
\small
{\baselineskip 8pt
\phantom{n}\verb+              t=5             +\\
\phantom{n}\verb+ .....................252..... +\\
\phantom{n}\verb+ ...................142....... +\\
\phantom{n}\verb+ .................152......... +\\
\phantom{n}\verb+ ...............252........... +\\
\phantom{n}\verb+ .............142............. +\\
\phantom{n}\verb+ ...........252............... +\\
{n}\verb+ .........142................. +\\
\phantom{n}\verb+ ........42................... +\\
\phantom{n}\verb+ ......43..................... +\\
\phantom{n}\verb+ ....54....................... +\\
\phantom{n}\verb+ ..53......................... +\\
\phantom{n}\verb+ 43........................... +\\
\phantom{n}\verb+               j              +\\
}
\end{minipage}
\begin{minipage}[t]{7cm}
\small
{\baselineskip 8pt
\phantom{n}\verb+              t=10             +\\
\phantom{n}\verb+ .....................55...... +\\
\phantom{n}\verb+ ...................44........ +\\
\phantom{n}\verb+ .................45.......... +\\
\phantom{n}\verb+ ...............55............ +\\
\phantom{n}\verb+ .............44.............. +\\
\phantom{n}\verb+ ...........55................ +\\
{n}\verb+ .........44.................. +\\
\phantom{n}\verb+ .......34.................... +\\
\phantom{n}\verb+ .....341..................... +\\
\phantom{n}\verb+ ...352....................... +\\
\phantom{n}\verb+ .351......................... +\\
\phantom{n}\verb+ 41........................... +\\
\phantom{n}\verb+               j              +\\
}
\end{minipage}
\end{center}
\end{figure}
\pagebreak
\begin{figure}[htbp]
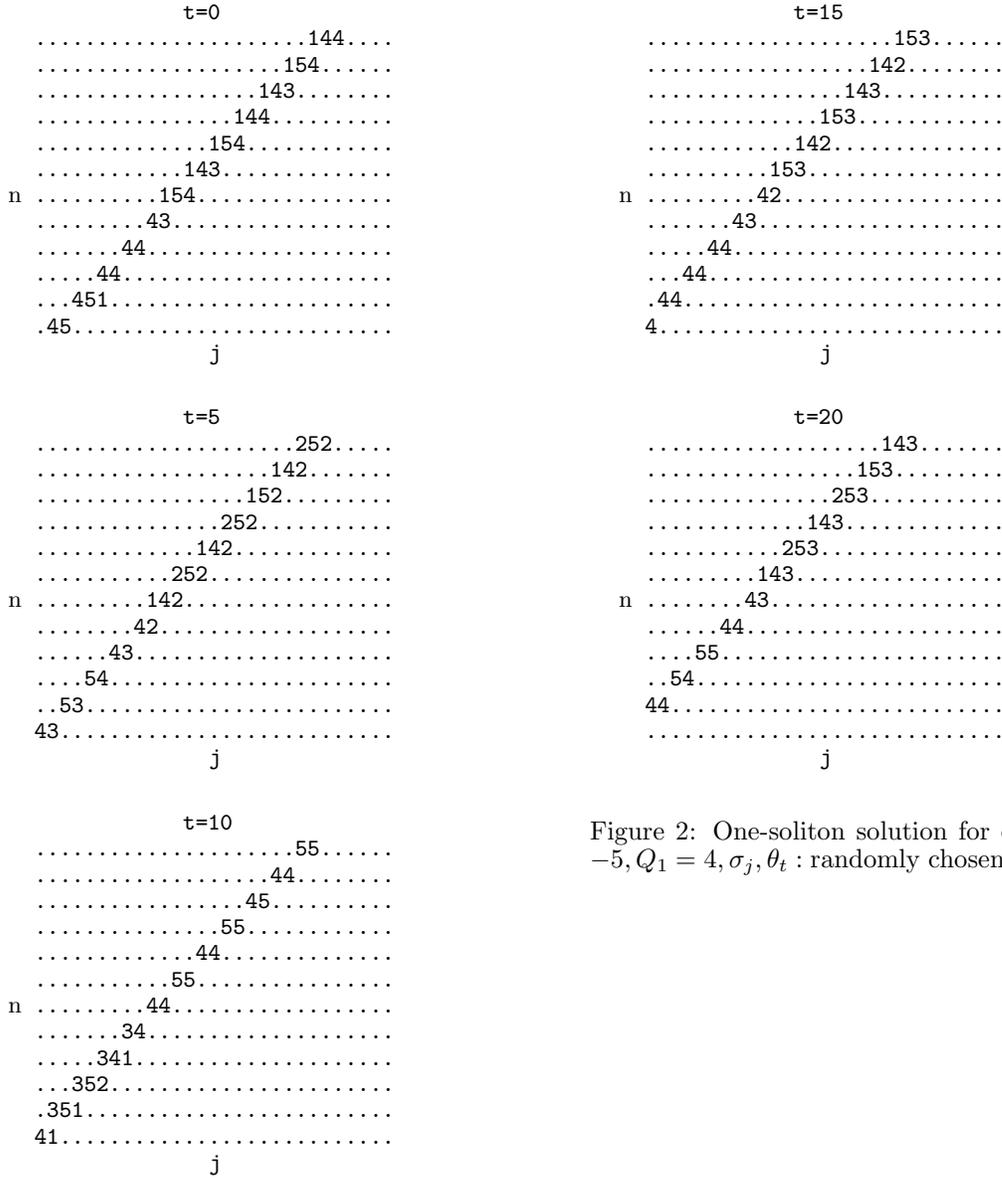

\begin{center}
\begin{minipage}[t]{7cm}
\small
{\baselineskip 8pt
\phantom{n}\verb+              t=15             +\\
\phantom{n}\verb+ ....................153...... +\\
\phantom{n}\verb+ ..................142........ +\\
\phantom{n}\verb+ ................143.......... +\\
\phantom{n}\verb+ ..............153............ +\\
\phantom{n}\verb+ ............142.............. +\\
\phantom{n}\verb+ ..........153................ +\\
{n}\verb+ .........42.................. +\\
\phantom{n}\verb+ .......43.................... +\\
\phantom{n}\verb+ .....44...................... +\\
\phantom{n}\verb+ ...44........................ +\\
\phantom{n}\verb+ .44.......................... +\\
\phantom{n}\verb+ 4............................ +\\
\phantom{n}\verb+               j              +\\
}
\end{minipage}
\begin{minipage}[t]{7cm}
\small
{\baselineskip 8pt
\phantom{n}\verb+              t=20             +\\
\phantom{n}\verb+ ...................143....... +\\
\phantom{n}\verb+ .................153......... +\\
\phantom{n}\verb+ ...............253........... +\\
\phantom{n}\verb+ .............143............. +\\
\phantom{n}\verb+ ...........253............... +\\
\phantom{n}\verb+ .........143................. +\\
{n}\verb+ ........43................... +\\
\phantom{n}\verb+ ......44..................... +\\
\phantom{n}\verb+ ....55....................... +\\
\phantom{n}\verb+ ..54......................... +\\
\phantom{n}\verb+ 44........................... +\\
\phantom{n}\verb+ ............................. +\\
\phantom{n}\verb+               j              +\\
}
\end{minipage}
\caption{One-soliton solution for eq.~\eqref{rutoda} ($P_1 = -5, 
Q_1 = 4, \sigma_j,\theta_t : \mbox{randomly chosen}$).}
\end{center}
\end{figure}
\clearpage

\begin{figure}[htbp]
\begin{center}
\begin{minipage}[t]{7cm}
\small
{\baselineskip 8pt
\phantom{n}\verb+               t=0             +\\
\phantom{n}\verb+ ............................. +\\
\phantom{n}\verb+ ...........................45 +\\
\phantom{n}\verb+ ........................154.. +\\
\phantom{n}\verb+ ......................253.... +\\
\phantom{n}\verb+ ....................352...... +\\
\phantom{n}\verb+ ..................451.....123 +\\
\phantom{n}\verb+ ...............154...12344321 +\\
\phantom{n}\verb+ ...........1235312344321..... +\\
{n}\verb+ ......1234434544321.......... +\\
\phantom{n}\verb+ .12344321.3521............... +\\
\phantom{n}\verb+ 4321....451.................. +\\
\phantom{n}\verb+ .....154..................... +\\
\phantom{n}\verb+ ...253....................... +\\
\phantom{n}\verb+ .352......................... +\\
\phantom{n}\verb+ 51........................... +\\
\phantom{n}\verb+ ............................. +\\
\phantom{n}\verb+               j              +\\
}
\end{minipage}
\begin{minipage}[t]{7cm}
\small
{\baselineskip 8pt
\phantom{n}\verb+               t=5             +\\
\phantom{n}\verb+ ............................4 +\\
\phantom{n}\verb+ .........................154. +\\
\phantom{n}\verb+ .......................253... +\\
\phantom{n}\verb+ .....................352..... +\\
\phantom{n}\verb+ ...................451....... +\\
\phantom{n}\verb+ ................154.......123 +\\
\phantom{n}\verb+ ..............253....12344321 +\\
\phantom{n}\verb+ ...........1352.12344321..... +\\
{n}\verb+ ......1234453344321.......... +\\
\phantom{n}\verb+ .1234432145321............... +\\
\phantom{n}\verb+ 4321..154.................... +\\
\phantom{n}\verb+ ....253...................... +\\
\phantom{n}\verb+ ..352........................ +\\
\phantom{n}\verb+ 451.......................... +\\
\phantom{n}\verb+ ............................. +\\
\phantom{n}\verb+ ............................. +\\
\phantom{n}\verb+               j              +\\
}
\end{minipage}
\end{center}
\end{figure}
\begin{figure}[htbp]
\begin{center}
\begin{minipage}[t]{7cm}
\small
{\baselineskip 8pt
\phantom{n}\verb+              t=10             +\\
\phantom{n}\verb+ ..........................154 +\\
\phantom{n}\verb+ ........................253.. +\\
\phantom{n}\verb+ ......................352.... +\\
\phantom{n}\verb+ ....................451...... +\\
\phantom{n}\verb+ .................154......... +\\
\phantom{n}\verb+ ...............253........123 +\\
\phantom{n}\verb+ .............352.....12344321 +\\
\phantom{n}\verb+ ...........451..12344321..... +\\
{n}\verb+ ......1235412344321.......... +\\
\phantom{n}\verb+ .1234433544321............... +\\
\phantom{n}\verb+ 4321.2531.................... +\\
\phantom{n}\verb+ ...352....................... +\\
\phantom{n}\verb+ .451......................... +\\
\phantom{n}\verb+ 4............................ +\\
\phantom{n}\verb+ ............................. +\\
\phantom{n}\verb+ ............................. +\\
\phantom{n}\verb+               j              +\\
}
\end{minipage}
\begin{minipage}[t]{7cm}
\small
{\baselineskip 8pt
\phantom{n}\verb+              t=15             +\\
\phantom{n}\verb+ .........................253. +\\
\phantom{n}\verb+ .......................352... +\\
\phantom{n}\verb+ .....................451..... +\\
\phantom{n}\verb+ ..................154........ +\\
\phantom{n}\verb+ ................253.......... +\\
\phantom{n}\verb+ ..............352.........123 +\\
\phantom{n}\verb+ ............451......12344321 +\\
\phantom{n}\verb+ .........154....12344321..... +\\
{n}\verb+ ......1253.12344321.......... +\\
\phantom{n}\verb+ .1234454344321............... +\\
\phantom{n}\verb+ 432135321.................... +\\
\phantom{n}\verb+ ..451........................ +\\
\phantom{n}\verb+ 54........................... +\\
\phantom{n}\verb+ ............................. +\\
\phantom{n}\verb+ ............................. +\\
\phantom{n}\verb+ ............................. +\\
\phantom{n}\verb+               j              +\\
}
\end{minipage}
\end{center}
\caption{Two-soliton solution for eq.~\eqref{rutoda} ($P_1 = -5,P_2=-2,
Q_1 = 4, Q_2 =3, \sigma_j = 1, \theta_t = 3$)}
\end{figure}

\begin{figure}[htbp]
\begin{center}
\begin{minipage}[t]{7cm}
\small
{\baselineskip 8pt
\phantom{n}\verb+               t=0             +\\
\phantom{n}\verb+ ............................. +\\
\phantom{n}\verb+ ............................. +\\
\phantom{n}\verb+ ............................4 +\\
\phantom{n}\verb+ .........................252. +\\
\phantom{n}\verb+ .......................341... +\\
\phantom{n}\verb+ ....................263...... +\\
\phantom{n}\verb+ ..................24......... +\\
{n}\verb+ ................34.......2223 +\\
\phantom{n}\verb+ .............133..133333111.. +\\
\phantom{n}\verb+ ..........23241332........... +\\
\phantom{n}\verb+ ..11233331.522............... +\\
\phantom{n}\verb+ 3221....141.................. +\\
\phantom{n}\verb+ ......242.................... +\\
\phantom{n}\verb+ ....441...................... +\\
\phantom{n}\verb+ .262......................... +\\
\phantom{n}\verb+               j              +\\
}
\end{minipage}
\begin{minipage}[t]{7cm}
\small
{\baselineskip 8pt
\phantom{n}\verb+               t=5             +\\
\phantom{n}\verb+ ............................. +\\
\phantom{n}\verb+ ............................5 +\\
\phantom{n}\verb+ .........................361. +\\
\phantom{n}\verb+ ......................143.... +\\
\phantom{n}\verb+ ....................352...... +\\
\phantom{n}\verb+ ..................33......... +\\
\phantom{n}\verb+ ................43.......2223 +\\
{n}\verb+ .............142..133333111.. +\\
\phantom{n}\verb+ ..........23331332........... +\\
\phantom{n}\verb+ ..112333311522............... +\\
\phantom{n}\verb+ 3221....24................... +\\
\phantom{n}\verb+ ......341.................... +\\
\phantom{n}\verb+ ...153....................... +\\
\phantom{n}\verb+ .351......................... +\\
\phantom{n}\verb+ 3............................ +\\
\phantom{n}\verb+               j              +\\
}
\end{minipage}
\end{center}
\end{figure}
\begin{figure}[htbp]
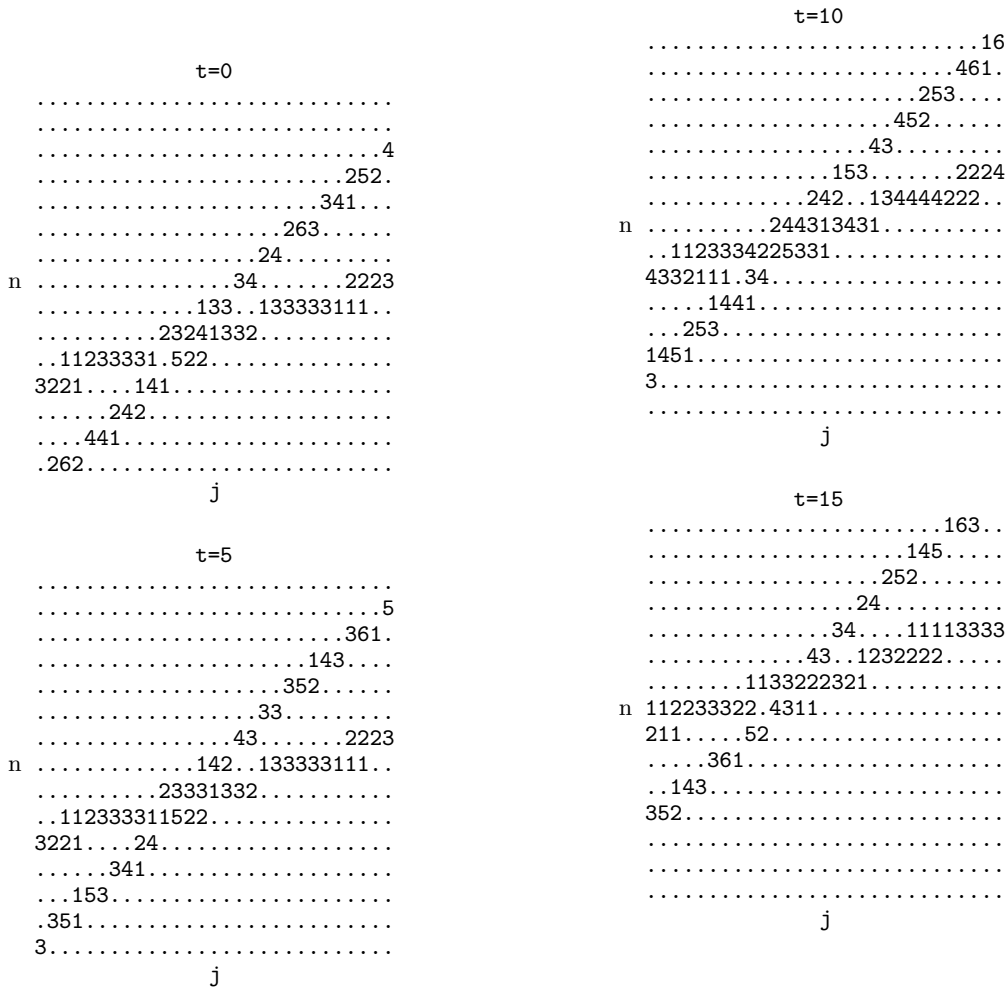

\begin{center}
\begin{minipage}[t]{7cm}
\small
{\baselineskip 8pt
\phantom{n}\verb+              t=10             +\\
\phantom{n}\verb+ ...........................16 +\\
\phantom{n}\verb+ .........................461. +\\
\phantom{n}\verb+ ......................253.... +\\
\phantom{n}\verb+ ....................452...... +\\
\phantom{n}\verb+ ..................43......... +\\
\phantom{n}\verb+ ...............153.......2224 +\\
\phantom{n}\verb+ .............242..134444222.. +\\
{n}\verb+ ..........244313431.......... +\\
\phantom{n}\verb+ ..1123334225331.............. +\\
\phantom{n}\verb+ 4332111.34................... +\\
\phantom{n}\verb+ .....1441.................... +\\
\phantom{n}\verb+ ...253....................... +\\
\phantom{n}\verb+ 1451......................... +\\
\phantom{n}\verb+ 3............................ +\\
\phantom{n}\verb+ ............................. +\\
\phantom{n}\verb+               j              +\\
}
\end{minipage}
\begin{minipage}[t]{7cm}
\small
{\baselineskip 8pt
\phantom{n}\verb+              t=15             +\\
\phantom{n}\verb+ ........................163.. +\\
\phantom{n}\verb+ .....................145..... +\\
\phantom{n}\verb+ ...................252....... +\\
\phantom{n}\verb+ .................24.......... +\\
\phantom{n}\verb+ ...............34....11113333 +\\
\phantom{n}\verb+ .............43..1232222..... +\\
\phantom{n}\verb+ ........1133222321........... +\\
{n}\verb+ 112233322.4311............... +\\
\phantom{n}\verb+ 211.....52................... +\\
\phantom{n}\verb+ .....361..................... +\\
\phantom{n}\verb+ ..143........................ +\\
\phantom{n}\verb+ 352.......................... +\\
\phantom{n}\verb+ ............................. +\\
\phantom{n}\verb+ ............................. +\\
\phantom{n}\verb+ ............................. +\\
\phantom{n}\verb+               j              +\\
}
\end{minipage}
\end{center}
\caption{Two-soliton solution for eq.~\eqref{rutoda} ($P_1 = -5,P_2=-2,
Q_1 = 4, Q_2 =3, \sigma_j, \theta_t : \mbox{\rm randomly chosen}$)}
\end{figure}

\onecolumn

\def\baselinestretch{1.2}

\section{Concluding Remarks}
We have presented a deautonomized 2D Toda lattice equation and its
ultra-discrete analogue. We have also found soliton solutions for the
latter case. The deautonomization resulting in eq.~\eqref{rutoda}
causes arbitrariness in the values of the soliton solutions, a property
which is not observed in the autonomous case. Deautonomization and
ultra-discretization of discrete soliton equations are expected to
find more applications in the field of computer science such as
convergence acceleration, interpolation and sorting algorithms.

\section*{Acknowledgements}
It is a pleasure to thank Professor Ryogo Hirota of Waseda University
for fruitful discussions, especially on deautonomized soliton
equations. One of the authors(R.W.) is a postdoctoral fellow of the
Fund for Scientific Research(F. W. O.), Flanders (Belgium). He would
also like to acknowledge the support of the F. W. O. through a
mobility grant. The present work is partially supported by a
Grant-in-Aid from the Japan Ministry of Education, Science and
Culture.

\end{document}